\documentclass[conference, review]{IEEEtran}

\usepackage{graphicx}
\usepackage[english]{babel}
\usepackage{lipsum}
\usepackage{cite}
\usepackage{xcolor}
\usepackage{dblfloatfix}
\usepackage{multirow}
\usepackage[para]{threeparttable}
\usepackage{balance}
\usepackage{import}
\usepackage{hyperref}
\usepackage{enumitem}
\usepackage{color}
\usepackage{booktabs}
\usepackage{bm}
\usepackage[caption=false]{subfig}
\usepackage{adjustbox}
\usepackage{amsmath}
\usepackage{relsize}
\usepackage[ruled,vlined,linesnumbered]{algorithm2e}
\usepackage{array}
\usepackage{caption}
\usepackage{subcaption}

\definecolor{darkgreen}{rgb}{0.4, 0.6, 0.3}
\definecolor{darkgray}{rgb}{0.6, 0.6, 0.6}

\usepackage{listings}
\definecolor{codegreen}{rgb}{0,0.6,0}
\definecolor{codegray}{rgb}{0.5,0.5,0.5}
\definecolor{codepurple}{rgb}{0.58,0,0.82}
\definecolor{backcolour}{rgb}{0.95,0.95,0.95}
\definecolor{black}{rgb}{0,0,0}

\lstdefinestyle{mystyle}{
    backgroundcolor=\color{backcolour},
    commentstyle=\color{codegreen},
    keywordstyle=\color{magenta},
    numberstyle=\tiny\color{codegray},
    stringstyle=\color{codepurple},
    basicstyle=\ttfamily\footnotesize,
    breakatwhitespace=false,
    breaklines=true,
    captionpos=b,
    keepspaces=true,
    numbers=left,
    numbersep=5pt,
    showspaces=false,
    showstringspaces=false,
    showtabs=false,
    tabsize=2
}

\lstdefinestyle{mystyle2}{
    backgroundcolor=\color{backcolour},
    commentstyle=\color{codegreen},
    keywordstyle=\color{black},
    numberstyle=\tiny\color{codegray},
    stringstyle=\color{codepurple},
    basicstyle=\ttfamily\footnotesize,
    breakatwhitespace=false,
    breaklines=true,
    captionpos=b,
    keepspaces=true,
    numbers=left,
    numbersep=5pt,
    showspaces=false,
    showstringspaces=false,
    showtabs=false,
    tabsize=2
}

\SetKw{Continue}{continue}

\ifCLASSINFOpdf
\else
\fi

\AtBeginDocument{\bstctlcite{BSTcontrol}}

\makeatletter
\def\@IEEEauthorblockconfadjspace{-0.5cm}
\makeatother

\begin{document}
\bstctlcite{BSTcontrol}

\title{FQTree: Fine-grained Quantization and Hardware Generation of Boosted Decision Trees}


\author{
    \IEEEauthorblockN{
        Zhiqiang Que\IEEEauthorrefmark{1}\IEEEauthorrefmark{2}\IEEEauthorrefmark{5},
        Chang Sun\IEEEauthorrefmark{1}\IEEEauthorrefmark{3},
        Haiyang Wang\IEEEauthorrefmark{3},
        Dinesh Pamunuwa\IEEEauthorrefmark{2},
        Roshan Weerasekera\IEEEauthorrefmark{2}, \\
        Qijia Tang\IEEEauthorrefmark{2},
        Bakhtiar Zadeh\IEEEauthorrefmark{4},
        Wayne Luk\IEEEauthorrefmark{4},
        Maria Spiropulu\IEEEauthorrefmark{3}
    }


    \IEEEauthorblockA{
        \IEEEauthorrefmark{2}
        University of Bristol, UK;
        \IEEEauthorrefmark{3}
        California Institute of Technology, USA;
        \IEEEauthorrefmark{4}
        Imperial College London, UK.
    }
    \vspace{-1cm}
}

\newcommand{\fmax}{$\mathrm{F}_\mathrm{max}$}

\maketitle

\begingroup\renewcommand\thefootnote{\IEEEauthorrefmark{1}}
\footnotetext{
Equal contribution 
\hspace{0.1em} \IEEEauthorrefmark{5} Corresponding author. Email: z.que@bristol.ac.uk
}
\endgroup


\begin{abstract}
    
Boosted decision trees (BDTs) are widely used in latency-critical applications, but efficient hardware deployment remains challenging. Existing designs often rely on uniform or manually tuned fixed-point formats, which can introduce unnecessary hardware cost or accuracy loss. This work presents the FQTree algorithm\footnote{\url{https://github.com/ecs-bristol/FQTree}} for fine-grained quantization-aware training of BDTs, together with the QXGB framework for automatic hardware generation. FQTree introduces a hardware-oriented leaf-value quantization scheme that uses a global quantization step together with a tree-wise shift, enabling compact non-negative integer leaf representations, controlled clipping/pruning, and bias folding to reduce datapath cost. This work further applies this quantization during boosting so that later trees adapt to the errors of the already-quantized ensemble, and then lowers the trained model into low-latency hardware implementations through a compiler-based flow. Results on JSC, MNIST, and NID show that our method reduces LUT usage by 26–57\% compared with the state-of-the-art FPGA-based BDT designs while matching or improving accuracy.

\end{abstract}

\IEEEpeerreviewmaketitle

\section{Introduction}

Boosted decision trees (BDTs) remain one of the most widely used machine learning models in real-world applications due to their strong predictive performance, compact model size, and relatively simple inference procedure~\cite{radovic2018machine}. Compared with large neural networks, BDTs are often easier to train, more robust on tabular data, and attractive for latency-critical scenarios where fast and deterministic decision making is required~\cite{radovic2018machine}. These properties make them appealing candidates for deployment on reconfigurable hardware platforms such as field-programmable gate arrays (FPGAs)~\cite{gajjar2022faxid, wang2022hardgbm, martinek2024lgbm2vhdl}, particularly in domains where strict latency and resource constraints are imposed.

Efficient FPGA deployment of BDTs, however, is still challenging. Although tree inference is structurally simpler than dense neural network inference, practical implementations must still balance prediction quality against hardware cost, including comparator width, storage footprint, accumulation precision, and overall datapath complexity. Existing FPGA-oriented BDT implementations typically rely on uniform or manually selected fixed-point representations across the model~\cite{conifer, treelut, ibdt}. Such a design choice is convenient, but it fails to reflect the heterogeneous numerical roles of different components in the ensemble. In a BDT, input features, split thresholds, leaf values, and the ensemble accumulation logic have very different sensitivity to quantization, and treating them with a single precision often leads either to unnecessary hardware cost or to avoidable accuracy degradation.

This challenge is especially acute for BDTs because quantization affects not only arithmetic precision, but also the routing decisions of the trees themselves. Small perturbations to features or thresholds may change branch outcomes and therefore alter the selected leaf entirely. As a result, straightforward post-training quantization is often not robust, particularly at low precision. This motivates the need for a quantization-aware training (QAT) approach that exposes the model to quantization effects during optimization, allowing it to adapt its numerical parameters to reduced precision while preserving predictive quality.

To address this challenge, this work presents a QAT algorithm for BDTs that adapts the precision of multiple numerical components of the model under hardware constraints. In particular, we observe that the resource cost of a BDT is dominated by the precision of leaf values, and therefore propose a magnitude-based precision assignment strategy integrated into training. Unlike post-training quantization, the bitwidths are not assigned after the model is fixed. Instead, during QAT, leaf values are learned under quantization and hardware-aware constraints, and their evolving magnitudes are used to guide precision allocation. Since BDT leaf values contribute additively to the output logits, their magnitude provides a direct indicator of their relative importance to the final prediction. As a result, the learned precision allocation naturally adapts to the importance of different leaves for classification accuracy. For instance, the early trees that make larger contributions receive more bits, while later trees that provide small corrective updates are represented with fewer bits. This allows the model to adapt its quantized parameters under quantization constraints, with the resulting precision assignments reflecting both task loss and hardware cost.

Beyond model optimization, we also target practical hardware realization. The quantized BDT produced by our training flow is automatically lowered to synthesizable hardware, enabling low-latency FPGA implementations without manual redesign of a separate datapath for each precision configuration. In this way, the proposed approach provides an end-to-end path from fine-grained QAT to hardware generation, which permits systematic exploration of the trade-off between accuracy, latency, and resource usage.

To the best of our knowledge, the FQTree algorithm together with the QXGB framework provides the first unified workflow for FPGA deployment of BDTs that combines hardware-oriented leaf-value quantization, quantization-aware training, and automatic hardware generation.

\noindent 
The main contributions of this work are:
\begin{itemize}[leftmargin=18pt] 
    \item \noindent The FQTree algorithm, a hardware-aware and fine-grained quantization-aware training approach for boosted decision trees, featuring a hardware-oriented leaf-value quantization formulation with a global quantization step, tree-wise shift, and bias folding. 
    \item \noindent The QXGB\footnote{\url{https://github.com/calad0i/qxgb}} framework (Quantized XGBoost, \texttt{qxgb}), a compiler-based scalable flow for generating efficient low-latency FPGA-based BDT implementations for both high-level synthesis (HLS) and RTL design, with a focus on minimizing latency and resource usage.

    \item \noindent A comprehensive evaluation of our method, showing that it reduces LUT usage by 26-57\% compared to the state-of-the-art BDT designs while matching or improving accuracy and achieving lower latency.
\end{itemize}

\section{Background and related work}
\label{sec:background}

A BDT model consists of an ensemble of $T$ decision trees whose outputs are combined additively. For an input feature vector $x$, the output is
\[
    \hat{y}(x)=\sum_{t=1}^{T}\eta_t f_t(x),
\]
where $f_t(x)$ is the output of tree $t$ and $\eta_t$ is its ensemble weight. Within each tree, an internal node compares a selected feature $x_{j_n}$ with a threshold $\theta_n$ and routes the input accordingly. Once a leaf node $l$ is reached, its leaf value $v_l$ is returned. The final prediction is obtained by accumulating the outputs of all trees.

BDTs are widely used for tabular data and have long supported classification and regression in high-energy physics~\cite{radovic2018machine}. Their compact models and efficient inference also make them attractive for latency-critical FPGA systems, such as real-time triggers.
Early work studied decision trees and random forests for scalable or high-throughput FPGA inference~\cite{kulaga2014fpga, owaida2017scalable, owaida2019distributed}. Later work extended this to Gradient BDTs (GBDTs), including memory-based and HLS-based accelerators, such as FAXID~\cite{gajjar2022faxid}, HardGBM~\cite{wang2022hardgbm}, and LGBM2VHDL~\cite{martinek2024lgbm2vhdl}. 
However, these works mainly focus on model export, compression, or architectural mapping rather than fine-grained quantization-aware optimization.
Existing quantized GBDT methods, such as Shi et al.~\cite{shi2022quantized},
primarily quantize gradients to accelerate training. By contrast, our work targets FPGA inference deployment and introduces a fine-grained QAT approach with automatic hardware generation.

Closely related to this work, Summers et al.\ introduce BDT support in Conifer to enable fully on-chip FPGA implementations with low latency \cite{conifer}. However, they only use uniform post-training quantization, where aggressive bit-width reduction can noticeably degrade accuracy. Alsharari et al.\ later propose QAT for GBDTs with integer-only and binary inference \cite{ibdt}.
More recently, TreeLUT~\cite{treelut} proposes a direct RTL generation flow with efficient LUT-based architectures and lightweight quantization. Compared with these works, our method targets a different point in the design space by introducing fine-grained QAT with separate precision control over features/thresholds and leaf values and demonstrating its benefits for hardware efficiency and accuracy. 

Fine-grained, hardware-aware quantization has also been explored for neural networks. HGQ learns per-parameter bitwidths for arbitrary-precision FPGA datapaths~\cite{hgq} and enables mixed-precision and low latency designs such as the GNN-based JEDI-linear~\cite{jedi-linear} and MLP-Mixer~\cite{mlpm}. HGQ-LUT further combines heterogeneous quantization with LUT-aware cost modeling and end-to-end compilation~\cite{sun2026hgq}. In contrast, FQTree addresses BDT-specific challenges.

\section{Challenges and Motivations}
\label{sec:design}
\subsection{Quantization challenges in BDTs}

Despite their apparent simplicity, efficient BDT hardware implementation remains challenging. Existing approaches commonly use a uniform bitwidth or a few manually selected precisions~\cite{conifer}, although different BDT quantities have distinct numerical roles and quantization sensitivities~\cite{ibdt}. Input features and split thresholds affect comparison accuracy and routing decisions, leaf values determine tree contributions, and accumulation requires sufficient dynamic range to prevent overflow. A single global precision may therefore waste bits on robust quantities while providing insufficient precision for sensitive ones.

The problem is further complicated by the discrete nature of tree inference. In neural networks, quantization typically perturbs arithmetic operations continuously. In contrast, for BDTs, quantization of a feature or threshold may flip a comparison result and redirect the input to a different subtree, leading to a discontinuous change in the output. This makes BDTs particularly sensitive to naive low-precision conversion and limits the effectiveness of straightforward post-training quantization.

\begin{figure*}
    \centering
    \includegraphics[width=0.76\textwidth]{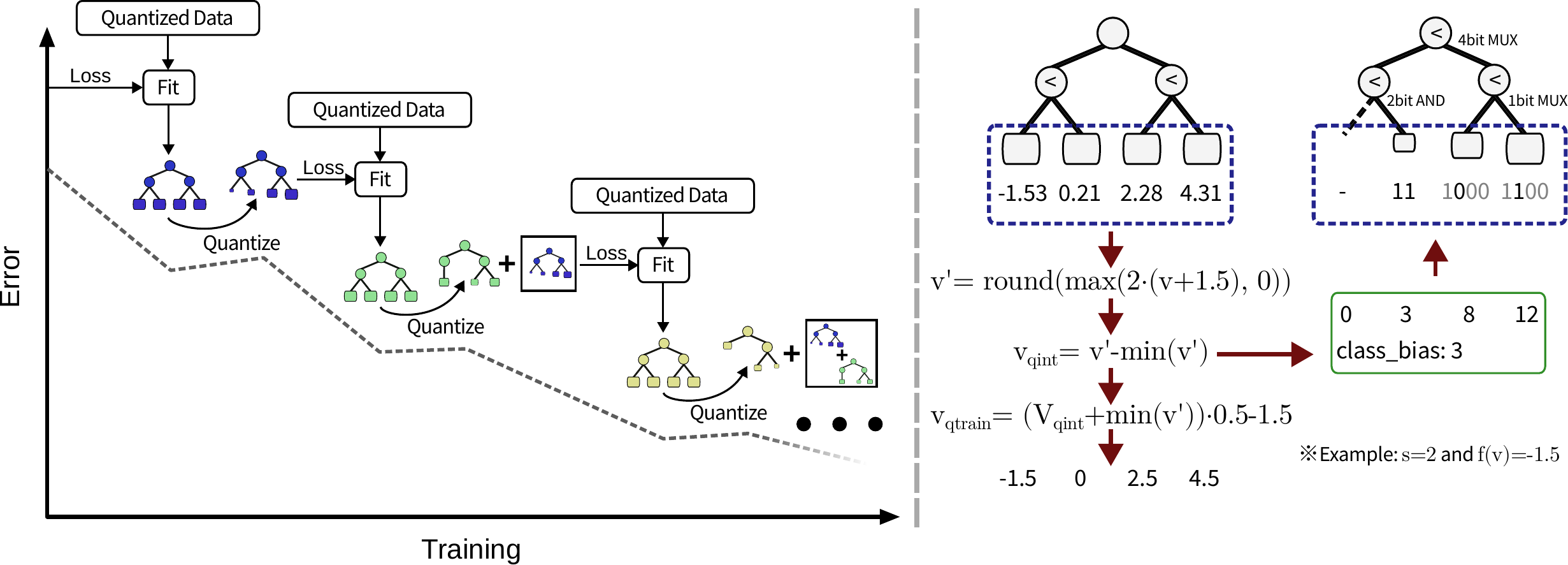}
    \caption{
        Illustration of the proposed FQTree algorithm. Leaf values are trained under quantization, with the resulting bitwidths determined directly by the magnitude of the quantized values. On the left, the dynamics of tree growth and quantization in a BDT ensemble are shown. On the right, we show an example of a single tree with heterogeneous leaf-value precision, where the number of bits required for muxing is smaller than the actual bits required to represent the leaf values.
    }
    \label{fig:qat}
    \vspace{-0.6cm}
\end{figure*}

\subsection{Why fine-grained quantization-aware training is needed}

These properties motivate a QAT approach rather than simple post-training quantization. By exposing the model to quantization effects during optimization, QAT allows the numerical parameters of the model to adapt to reduced precision. This is especially important for BDTs, where quantization influences not only arithmetic values but also the routing behavior of the ensemble.

Moreover, a fine-grained mixed-precision approach is desirable because the precision requirements of different model components are inherently heterogeneous. Features/thresholds and leaf values should not be treated identically. A more flexible approach is to optimize their precision separately, while constraining the search space through structured grouping so that the resulting model remains hardware-friendly. Such an approach enables a better trade-off between predictive accuracy and implementation cost than conventional uniform fixed-point baselines.

\subsection{Hardware-aware optimization for FPGA deployment}

For FPGA deployment, precision directly affects implementation cost. Narrower feature and threshold representations reduce comparator width and storage cost, while lower-precision leaf values simplify the datapath and reduce memory footprint. The precision of the ensemble adder tree also determines both adder cost and output range. Therefore, precision optimization should consider not only task loss, but also hardware cost.

This motivates an end-to-end flow in which the model is trained under quantization and then lowered automatically to hardware. Such a flow avoids repeated manual redesign when bitwidth assignments change and enables systematic exploration of the accuracy-latency-resource trade-off. Based on this idea, we develop the FQTree algorithm for fine-grained QAT of BDTs and use the QXGB framework for automated hardware generation for low-latency FPGA implementations.

\section{Methodology}
\subsection{Overview}

We propose the \textbf{FQTree} algorithm, a fine-grained QAT approach for BDTs targeting low-latency FPGA deployment. In contrast to conventional post-training fixed-point conversion, FQTree incorporates quantization into training so that the final model is optimized under the same numerical constraints that will be used in hardware. The tree structure and split thresholds are kept unchanged, while the leaf outputs are quantized in a hardware-friendly form.

Figure~\ref{fig:qat} illustrates the overall workflow. On the left, the BDT ensemble is trained in the usual stage-wise boosting manner, but each newly fitted tree is immediately quantized before being used by the subsequent training stages. As a result, later trees are trained on the residual errors produced by the already-quantized ensemble, rather than by an ideal floating-point one. This allows the subsequent boosting stages to compensate not only for task residuals, but also for the distortion introduced by quantization, so the final ensemble is better matched to the behavior of the deployed low-precision model. On the right, the figure shows how the leaf values of one tree are transformed into a compact non-negative integer representation and an associated bias term. This representation is then used for hardware generation, where smaller integer ranges translate into simpler multiplexing-accumulation logic.

\subsection{Leaf-value quantization}

Instead of assigning one uniform fixed-point format to the whole ensemble, FQTree quantizes each tree's leaf values according to their own numerical range. Empirically, earlier trees in the boosting sequence usually have a larger impact on the final prediction and often require a larger dynamic range, whereas later trees can typically be quantized more aggressively with little accuracy loss. Hence, instead of normalizing the values by min-max scaling at the tree level, we fix the step size for the leaf values globally across the ensemble. Since the classification results are determined by the relative order of the outputs invariant to scaling, we allow a global scaling factor during the training process. This factor is removed before hardware generation. For the leaf-value vector $\vec{v}$ of one tree, the quantization used during training is defined as
\begingroup
\setlength{\abovedisplayskip}{3pt}
\setlength{\abovedisplayshortskip}{1pt}
\setlength{\belowdisplayskip}{3pt}
\setlength{\belowdisplayshortskip}{3pt}
\begin{align}
    \vec{v}'
     & = \operatorname{round}\!\left(\max\!\left(s\cdot(\vec{v}-f(\vec{v})),\,0\right)\right), \label{eq:q1} \\
    \vec{v}_{\mathrm{qint}}
     & = \vec{v}'-\min(\vec{v}'), \label{eq:q2}                                                              \\
    \vec{v}_{\mathrm{qtrain}}
     & = \left(\vec{v}_{\mathrm{qint}}+\min(\vec{v}')\right)\cdot s^{-1}+f(\vec{v}), \label{eq:q3}
\end{align}
\endgroup
where $\vec{v}_{\mathrm{qint}}$ is the integer representation used for hardware generation, $\vec{v}_{\mathrm{qtrain}}$ is the dequantized value used in training, $s$ is the global step size, and $f(\vec{v})$ is a tree-wise shift factor.

Equation~\eqref{eq:q1} first applies a global quantization step and a tree-wise shift, then clips negative values to zero. Equation~\eqref{eq:q2} further rebases the integer outputs so that the smallest leaf value becomes zero.
Equation~\eqref{eq:q3} dequantizes the integer leaf values for use in ensemble prediction. The resulting quantized predictions are then used to compute the loss gradients or residuals for the subsequent boosting round.

The maximum effective bitwidth of one tree is determined by the dynamic range of $\vec{v}_{\mathrm{qint}}$, i.e.
\begingroup
\setlength{\abovedisplayskip}{3pt}
\setlength{\abovedisplayshortskip}{1pt}
\setlength{\belowdisplayskip}{3pt}
\setlength{\belowdisplayshortskip}{3pt}
\begin{equation}
    b_t = \left\lceil \log_2 \left(\max(\vec{v}_{\mathrm{qint}})+1\right) \right\rceil.
\end{equation}
\endgroup
Hence, trees with a smaller post-quantization range naturally require fewer bits in hardware.

\subsection{Feature/threshold quantization}

For feature and threshold quantization, we use the standard uniform quantization. In hardware, since the comparison is performed by a signed subtractor and the sign bit is used as the output, the precision of the feature and threshold is naturally aligned. If, after boosting, the algorithm selects a threshold of higher precision than the feature, the extra bits of the threshold will not contribute to the decision and will be rounded to the same precision as the feature.

\begin{figure*}
    \centering
    \includegraphics[width=0.76\textwidth]{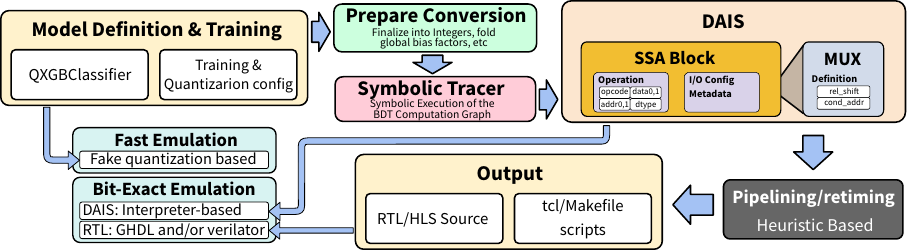}
    \caption{
        Overall workflow of our method. The FQTree algorithm provides fine-grained precision assignment during training, and the QXGB framework directly lowers the quantized BDT to synthesizable hardware. A key enabler is the extension of the DAIS IR in da4ml~\cite{da4ml} to represent both arithmetic and tree-based operations through the MUX operator, allowing quantized BDT inference to be captured in a unified static dataflow form. Once generated, this IR supports bit-exact emulation within seconds after training and serves as the basis for hardware generation.
    }
    \label{fig:workflow}
    \vspace{-0.5cm}
\end{figure*}

\subsection{Shift factor and bias folding}

The shift factor $f(\vec{v})$ serves two purposes. First, it pushes the leaf values toward a non-negative representation, which reduces the cost of the selection logic. For example, a set of leaf values such as $[-1,0,1,2]$ requires signed representation, whereas shifting it to $[0,1,2,3]$ removes the sign bit and simplifies the hardware datapath. Second, by allowing a controlled amount of clipping, the shift can increase the number of zeros in the quantized leaf vector, which effectively prunes small-magnitude outputs.
In practice, we use constant
\begingroup
\setlength{\abovedisplayskip}{3pt}
\setlength{\abovedisplayshortskip}{1pt}
\setlength{\belowdisplayskip}{3pt}
\setlength{\belowdisplayshortskip}{3pt}
\begin{equation}
    f(\vec{v}) =  b_{\max},
\end{equation}
\endgroup
where $b_{\max}$ is a tunable hyperparameter. If $b_{\max}$ is larger than the minimum leaf value of a tree, some small leaf values are clipped to zero after quantization. Therefore, $b_{\max}$ controls the trade-off between accuracy and sparsity in the leaf outputs.

The offset removed in Equation~\eqref{eq:q2} does not need to be carried through the per-leaf selection logic. Instead, it can be folded into a tree-level or class-level bias term and added only once after tree accumulation. From Equation~\eqref{eq:q3}, the quantized training value can be rewritten as
\begingroup
\setlength{\abovedisplayskip}{3pt}
\setlength{\abovedisplayshortskip}{1pt}
\setlength{\belowdisplayskip}{3pt}
\setlength{\belowdisplayshortskip}{3pt}
\begin{equation}
    \vec{v}_{\mathrm{qtrain}}
    = \vec{v}_{\mathrm{qint}}\cdot s^{-1}
    + \left(\min(\vec{v}')\cdot s^{-1} + f(\vec{v})\right).
\end{equation}
\endgroup
Thus, the per-tree offset is separated from the integer leaf representation and absorbed into the ensemble bias. In hardware, this is much cheaper than carrying signed offsets through every tree path.

\section{Hardware Generation}

\subsection{From Quantized BDTs to a Dataflow Representation}

In the QXGB framework, after quantization-aware training, the BDT model is exported with a fixed tree topology and explicit precision assignments for features/thresholds and leaf values. These quantized parameters are then passed to a static dataflow compiler, which automatically lowers the model into synthesizable RTL or HLS-ready hardware.
Cross-stage design-flow frameworks such as MetaML-Pro~\cite{metamlpro} automate the joint exploration of model transformations and hardware implementation parameters for efficient FPGA acceleration. This work builds on this general co-design philosophy but targets boosted decision trees. 

Figure~\ref{fig:workflow} illustrates the overall flow. The process starts from model definition and training, where the quantized BDT and its quantization configuration are obtained. A preparation step then finalizes the trained parameters into an integer representation, for example by folding global scaling factors and resolving the fixed-point metadata required by hardware generation. The compiler next symbolically traces the BDT computation graph and lowers it into an extended Distributed Arithmetic Instruction Set (DAIS)~\cite{da4ml} intermediate representation (IR). This representation is subsequently used for emulation, backend code generation, and implementation-oriented optimizations such as pipelining and retiming. Compared with a software-style implementation of tree traversal, this compilation flow treats BDT inference as a stateless dataflow kernel with deterministic latency and explicit fixed-point semantics, which is more suitable for low-latency FPGA deployment.

\subsection{Extending DAIS for Tree Inference}

Our implementation builds on the DAIS IR introduced in da4ml~\cite{da4ml}. DAIS was originally designed as a low-level SSA-based representation for arithmetic-dominated machine-learning workloads, especially distributed-arithmetic implementations of neural networks. BDT inference, however, is structurally different: instead of being dominated by multiply-accumulate style operators, it mainly consists of feature-threshold comparisons, conditional routing, and leaf-value selection.

To represent this computation pattern naturally, we extend DAIS with operations required by tree inference. The most important addition is an explicit \texttt{MUX} instruction, which captures the conditional selection behavior of decision nodes directly in the IR. This avoids forcing tree traversal into an unnatural arithmetic template. The extended IR therefore remains compact and SSA-based, while becoming expressive enough to represent both arithmetic datapaths and tree-structured decision logic in a unified compiler flow.

\subsection{Lowering quantized BDTs to hardware}

Given a trained quantized BDT, the compiler first extracts the structure of each tree, including the feature index and threshold of every internal node, the quantized values stored at the leaves, and the accumulation rule of the ensemble. Since all of this information is known at compile time, the model can be lowered into a fully static dataflow graph.

Each internal decision node is mapped to a fixed-point comparison between a quantized input feature and a quantized threshold. The resulting one-bit decision signal drives a \texttt{MUX} structure that selects between the outputs of the left and right subtrees. Applying this rule recursively transforms each tree into a comparator--mux network whose final output is the quantized value of the selected leaf. This lowering makes the hardware implications of fine-grained quantization explicit.

\subsection{Generated architecture}

The generated accelerator consists of four main stages. First, the input stage quantizes and aligns the input features according to the precisions determined by the trained model. Since different features may use different bitwidths, this stage is heterogeneous by construction. Second, the node-evaluation stage performs fixed-point comparisons for all required feature-threshold pairs and emits one-bit routing decisions. Third, the tree-evaluation stage uses these routing decisions to drive a hierarchy of \texttt{MUX} operations or, when beneficial, a lookup-based realization. In the mux-based realization, each internal node selects between the outputs of its left and right child subtrees, and the final output of the tree is the quantized value associated with the selected leaf. Finally, the ensemble stage accumulates the outputs of all trees using an adder tree. The precision of the adders at each branch of the accumulator tree is determined by the leaf value bitwidths being added, where the heterogeneous leaf-value quantization leads to narrower accumulators locally in the adder tree.

\subsection{Emulation and Backend Generation}

A practical advantage of the flow in Figure~\ref{fig:workflow} is that emulation is available immediately after lowering. During model development, fast emulation can be performed using fake-quantization-based execution. After DAIS generation, the flow also supports bit-exact emulation through either an interpreter over the DAIS representation or RTL-level simulation, for example using GHDL or Verilator. This allows the quantized model and the generated hardware to be validated before full implementation. Once the dataflow graph has been finalized, the backend emits platform-agnostic RTL or HLS-ready source code, together with the scripts needed for downstream builds. Additional backend optimizations, such as pipelining and retiming, can then be applied for the target device. Overall, this provides a direct path from fine-grained quantization-aware training to low-latency FPGA hardware generation for BDTs.

\section{Evaluation}

\subsection{Experimental Setup}

We evaluate our method on three datasets: the MNIST handwritten digit classification dataset~\cite{mnist}, the OpenML jet substructure classification (JSC) dataset~\cite{openml-jet} in high-energy physics, and a binarized version~\cite{unsw-nb15-bin} of the UNSW-NB15 network intrusion detection (NID) dataset~\cite{unsw-nb15}. These benchmarks are widely used in prior studies of ultra-low-latency and area-efficient tree-based hardware, and were chosen to cover different application domains and quantization settings.
The MNIST and NID datasets are quantized to binary features, and the JSC dataset is quantized to 8-bit fixed-point features. MNIST and JSC are multiclass classification problems with 10 and 5 classes, respectively, while NID is a binary classification problem. For all datasets, we use the standard training and test splits without additional data augmentation, and train the BDT models using XGBoost~\cite{xgboost}.

Unless otherwise stated, we use {xcvu13p-flga2577-2-e} as the target device for all experiments. We also use the out-of-context, post-routing reports from Vivado 2025.1 for resource and \fmax\ measurements. Verilator~\cite{verilator} is used for bit-and-cycle-accurate verification of the generated RTL code. Model accuracy is achieved on the RTL-based model.

\begin{figure}
    \begin{center}
        \includegraphics[width=0.7\linewidth]{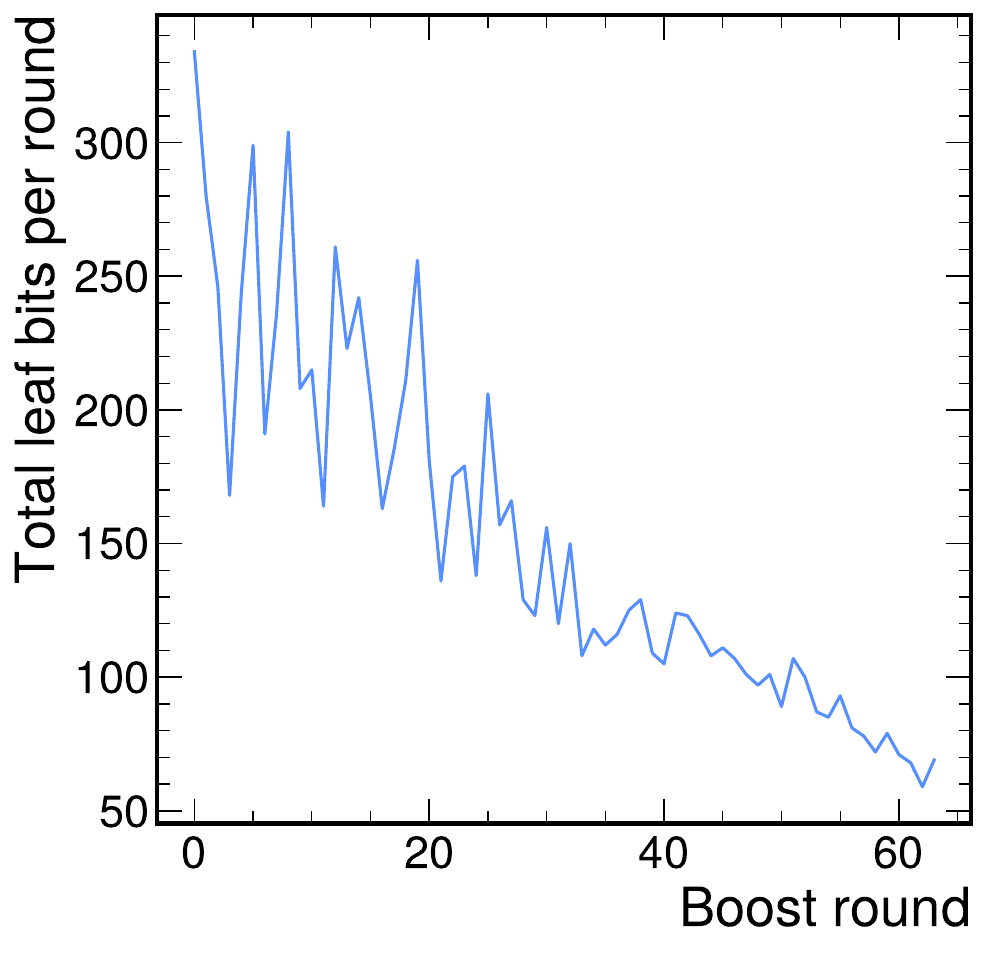}
    \end{center}
    \vspace{-0.5cm}
    \caption{Total number of bits assigned to the MNIST classifier BDT leaf values as a function of boost round index with a maximum tree depth 5. The bitwidth demand is highest for the first few trees and drops gradually for later trees, which is consistent with the boosting process where early trees make larger contributions to the final prediction.}
    \label{fig:bdt_bw}
    \vspace{-0.5cm}
\end{figure}

\subsection{Leaf-Value Bit Allocation Across Boosting Rounds}

Figure~\ref{fig:bdt_bw} shows how the total number of bits assigned to leaf values changes across boosting rounds for the MNIST classifier with maximum tree depth 5. In this model, each boosting round contains 10 trees, since one tree is constructed for each output class. Therefore, the value shown for each round is the sum of the leaf-value bitwidths over the 10 class-specific trees generated in that round.

Figure~\ref{fig:bdt_bw} also shows a clear overall decreasing trend. The earliest boosting rounds require substantially more bits, while later rounds gradually become less demanding in terms of leaf-value precision. This behavior is consistent with the standard boosting process: early trees usually capture the dominant decision patterns and therefore contribute more strongly to the final prediction, whereas later trees mainly refine residual errors and make smaller corrections. The result highlights a strong heterogeneity across trees in terms of numerical precision requirements, suggesting that a uniform bitwidth assignment is inefficient. Instead, it motivates a fine-grained quantization strategy that allocates more bits to the early trees and fewer bits to the later ones, thereby reducing hardware cost without affecting model accuracy.

\subsection{Accuracy-Resource Trade-offs with FQTree}

\begin{figure}
    \begin{center}
        \includegraphics[height=0.78\linewidth]{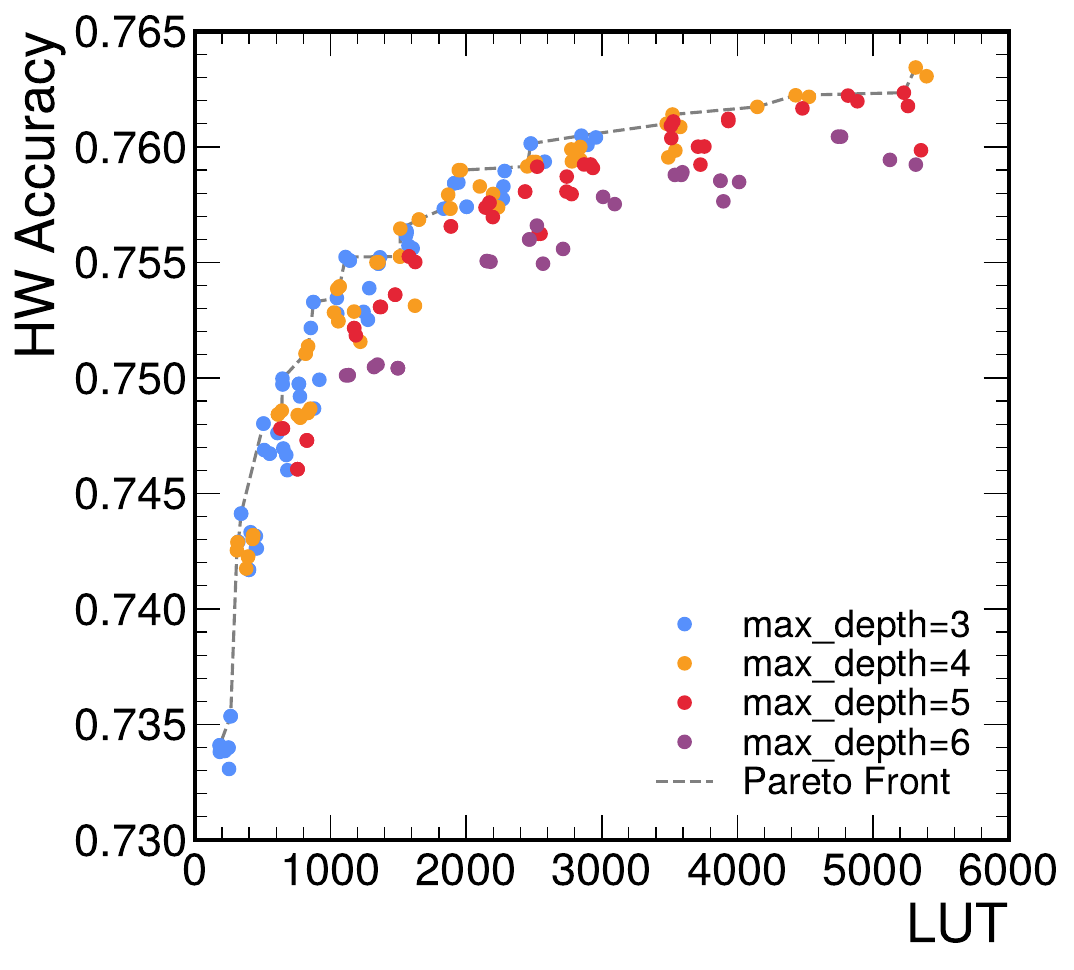}
    \end{center}
    \vspace{-0.5cm}
    \caption{Design space explored by FQTree for JSC using BDTs with maximum depth $3-6$. Each point represents one quantization configuration and shows the trade-off between LUT usage and hardware accuracy.
    }
    \label{fig:jsc}
    \vspace{-0.5cm}
\end{figure}

\begin{figure}
    \begin{center}
        \includegraphics[height=0.78\linewidth]{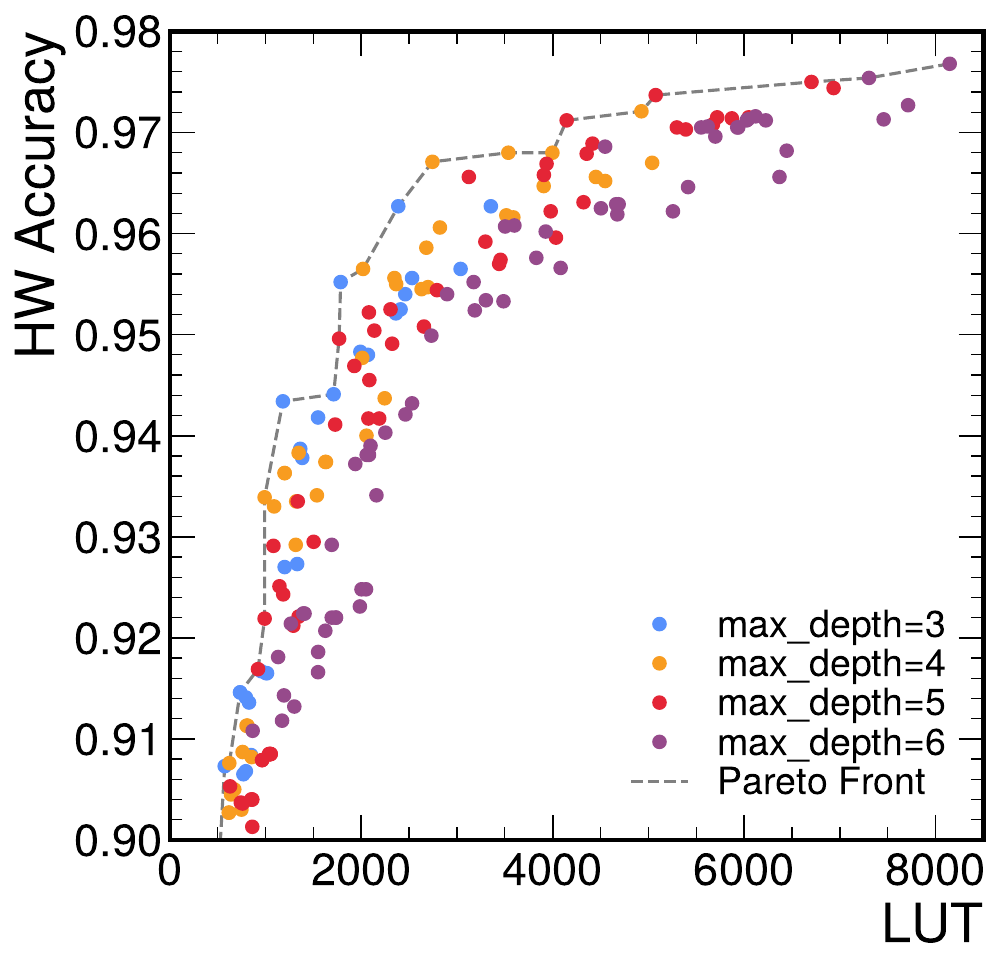}
    \end{center}
    \vspace{-0.5cm}
    \caption{Design space explored by FQTree for the MNIST dataset using BDTs with maximum depth $3-6$. Each point corresponds to one quantization configuration and illustrates the trade-off between LUT usage and hardware accuracy. As the tree depth increases, the achievable accuracy improves, accompanied by higher LUT consumption.}
    \label{fig:mnist}
    \vspace{-0.5cm}
\end{figure}

Figures~\ref{fig:jsc} and~\ref{fig:mnist} show the design spaces explored by FQTree for the JSC and MNIST datasets, respectively. In both cases, we sweep a wide range of quantization configurations for BDTs with different maximum depths and evaluate the resulting hardware implementations in terms of LUT usage and hardware accuracy. The figures show that FQTree exposes a rich set of implementation points spanning different resource budgets, rather than a single fixed design, making it possible to select operating points according to the desired balance between accuracy and hardware cost.

For JSC, Figure~\ref{fig:jsc} shows a clear accuracy--resource trade-off. As LUT usage increases, the hardware accuracy generally improves, although the gains gradually diminish in the high-resource region. The figure also suggests that moderate tree depths, especially \texttt{max\_{depth}}=4, provide the most favorable accuracy--resource trade-off, while deeper trees do not consistently improve accuracy despite their higher LUT cost. The Pareto frontier indicates that moving from small to moderate LUT budgets yields the largest accuracy gains, whereas further LUT increases provide only incremental improvements. This behavior suggests that FQTree can effectively identify efficient operating points for JSC under different hardware constraints.

A similar trend can be observed for MNIST in Figure~\ref{fig:mnist}, but with a wider LUT range and a higher overall accuracy range. Larger models and deeper trees tend to achieve better accuracy, while requiring more LUT resources. The Pareto frontier highlights that FQTree can systematically capture the best achievable trade-offs across quantization settings and tree depths.
Overall, these two figures demonstrate that our method provides a practical way to explore and select quantized BDT implementations that match different application-level accuracy targets and FPGA resource budgets.

\begin{table*}
    \newcommand{\nrow}[2]{\multirow{#1}{*}{#2}}

    \caption{
        Performance and resource comparison of BDT implementations on the JSC HLF, MNIST and NID datasets. FQTree achieves better accuracy-resource-latency trade-offs than prior BDT-based designs and conventional PTQ.
    }
    \label{tab:bdt}

        \begin{threeparttable}
    \begin{adjustbox}{width=0.86\textwidth,center=\textwidth}
            \begin{tabular}{ll|clccccc}
                \midrule
                Task              & Implementation                       & Acc. $\uparrow$ & Latency [cycles] & LUT       & DSP & FF     & \fmax (MHz)     & II \\
                \midrule
                \nrow{8}{JSC HLF} & \textbf{FQTree} (\textbf{This work}) & 75.7\%          & 2 (4.0 ns)      & 1,652     & 0   & 446    & 505      & 1  \\
                                  & \textbf{FQTree} (\textbf{This work}) & 74.8\%          & 1 (2.0 ns)      & 548       & 0   & 146    & 499      & 1  \\
                                  & PTQ (\textbf{This work})             & 75.6\%          & 2 (4.0 ns)      & 2,194     & 0   & 361    & 501      & 1  \\
                                  & PTQ (\textbf{This work})             & 74.7\%          & 1 (2.0 ns)      & 642       & 0   & 143    & 502      & 1  \\
                                  & FPGA'25~\cite{treelut} TreeLUT       & 75.6\%          & 3$^a$ (3.9 ns)  & 2,234     & 0   & 347    & 735      & 1  \\
                                  & FPGA'25~\cite{treelut} TreeLUT       & 74.6\%          & 2$^a$ (2.1 ns)  & 796       & 0   & 74     & 887      & 1  \\
                                  & TCAS-I'25~\cite{ibdt} QBDT-8bit      & 75\%$^c$       & 5 (7.1 ns)      & 6,500$^b$ & -   & -      & 670      & 1  \\
                                  & TCAS-I'25~\cite{ibdt} QBDT-1bit      & 73.5\%          & 1 (1.4 ns)      & 906       & -   & -      & 724      & 1  \\
                                  & JINST'20~\cite{conifer}  Conifer     & 73.8\%          & 12 (60 ns)      & 96,148    & -   & 42,802 & $\sim200$ & 1  \\ 

                \midrule
                \nrow{8}{MNIST}   & \textbf{FQTree} (\textbf{This work}) & 97.7\%          & 2 (4.0 ns)      & 8,147     & 0   & 2,873  & 504      & 1  \\
                                  & \textbf{FQTree} (\textbf{This work}) & 96.7\%          & 2 (3.5 ns)      & 2,744     & 0   & 1,156  & 576      & 1  \\
                                  & \textbf{FQTree} (\textbf{This work}) & 95.6\%          & 2 (3.2 ns)      & 2,019     & 0   & 1,150  & 633      & 1  \\
                                  & PTQ (\textbf{This work})             & 96.6\%          & 2 (3.9 ns)      & 4,439     & 0   & 1,374  & 518      & 1  \\
                                  & PTQ (\textbf{This work})             & 96.0\%          & 2 (3.4 ns)      & 3,377     & 0   & 1,478  & 584      & 1  \\
                                  & FPGA'25~\cite{treelut} TreeLUT       & 96.6\%          & 3$^a$ (3.6 ns)  & 4,478     & 0   & 597    & 791      & 1  \\
                                  & FPGA'25~\cite{treelut} TreeLUT       & 95.6\%          & 3$^a$ (3.2 ns)  & 3,499     & 0   & 759    & 874      & 1  \\
                                  & JSPS'20~\cite{polybinn} POLYBiNN    & 97.2\%          & 900 (90 ns)     & 109,653   & -   & -      & 100       & -  \\
                                  & JSPS'20~\cite{polybinn} POLYBiNN    & 95.6\%          & 700 (70 ns)     & 9,943     & -   & -      & 100       & -  \\

                \midrule
                \nrow{7}{NID}     & \textbf{FQTree} (\textbf{This work}) & 93.1\%          & 1 (1.9 ns)      & 157       & 0   & 155    & 524.      & 1  \\
                                  & \textbf{FQTree} (\textbf{This work}) & 92.8\%          & 1 (1.4 ns)      & 83        & 0   & 76     & 740      & 1  \\
                                  & \textbf{FQTree} (\textbf{This work}) & 91.7\%          & 1 (1.1 ns)      & 38        & 0   & 45     & 871      & 1  \\
                                  & FPGA'25~\cite{treelut} TreeLUT       & 92.7\%          & 2$^a$ (2.7 ns)  & 345       & 0   & 33     & 681      & 1  \\
                                  & FPGA'25~\cite{treelut} TreeLUT       & 91.5\%          & 2$^a$ (1.7 ns)  & 89        & 0   & 19     & 1047     & 1  \\
                                  & TCAS-I'25~\cite{ibdt} QBDT-8bit      & 92\%$^c$       & 5 (7.1 ns)      & 1,800$^b$ & -   & -      & 714      & 1  \\
                                  & TCAS-I'25~\cite{ibdt} QBDT-1bit      & 91.5\%          & 1 (1.4 ns)      & 170       & -   & -      & 724      & 1  \\

                \midrule
            \end{tabular}
    \end{adjustbox}
            \footnotesize
            \textbf{$^a$} TreeLUT reports cycles differently and uses unregistered I/O with manual external path delays; for fair comparison, we normalize its metrics to the same convention as the other implementations. \hspace{0.2cm} 
            \textbf{$^b$} The number of LUTs has only two significant digits, and the exact value is not provided in the original paper. \\
            \textbf{$^c$} The accuracy has only two significant digits, and the exact value is not provided in the original paper.\\
            \normalsize
            \vspace{-0.3cm}

        \end{threeparttable}
    \vspace{-0.5cm}
\end{table*}


\subsection{Performance Comparison and Discussion}

Table~\ref{tab:bdt} shows that FQTree achieves a better accuracy-resource-latency trade-off than prior FPGA-based studies as well as our PTQ baselines.

On the JSC HLF task, FQTree reaches the best accuracy, achieving 75.7\% with 1,652 LUTs and a latency of 2 cycles~(4.0 ns). Compared with TreeLUT~\cite{treelut}, which reports 75.6\% accuracy with 2,234 LUTs and 3 cycles, FQTree slightly improves accuracy while reducing LUT usage by about 26\% and shortening the pipeline depth. Although the absolute latency is very similar, FQTree achieves a better hardware efficiency. At a lower-cost point, FQTree achieves 74.8\% accuracy using only 548 LUTs and 1 cycle~(2.0 ns), compared with 74.6\% accuracy, 796 LUTs, and 2 cycles for TreeLUT. This corresponds to about 31\% fewer LUTs, with slightly better accuracy and lower latency. Compared with QBDT-8bit~\cite{ibdt}, which achieves 75\% accuracy using approximately 6,500 LUTs and 5 cycles, FQTree provides slightly higher accuracy while using far fewer resources and substantially lower latency. FQTree also compares very favourably with QBDT-1bit~\cite{ibdt} and Conifer~\cite{conifer}, both of which deliver lower accuracy and/or much higher hardware cost.

On MNIST, FQTree's highest-accuracy configuration reaches 97.7\% with 8,147 LUTs and 2 cycles (4.0 ns), outperforming the best previously reported accuracy of 97.2\% from POLYBiNN~\cite{polybinn}, while requiring far fewer LUTs and much lower latency. At a more moderate point, FQTree achieves 96.7\% accuracy with 2,744 LUTs and 2 cycles~(3.5 ns), whereas TreeLUT reports 96.6\% with 4,478 LUTs and 3 cycles. This gives FQTree a small accuracy improvement, about 39\% lower LUT usage, and slightly lower absolute latency. At another point, FQTree reaches 95.6\% with only 2,019 LUTs, compared with 95.6\% and 3,499 LUTs for TreeLUT, corresponding to about 42\% LUT reduction at the same latency. These results show that FQTree extends the Pareto frontier on MNIST, either improving accuracy at similar cost or reducing hardware cost substantially at comparable accuracy.

\begin{figure}
    \begin{center}
        \includegraphics[width=0.96\linewidth]{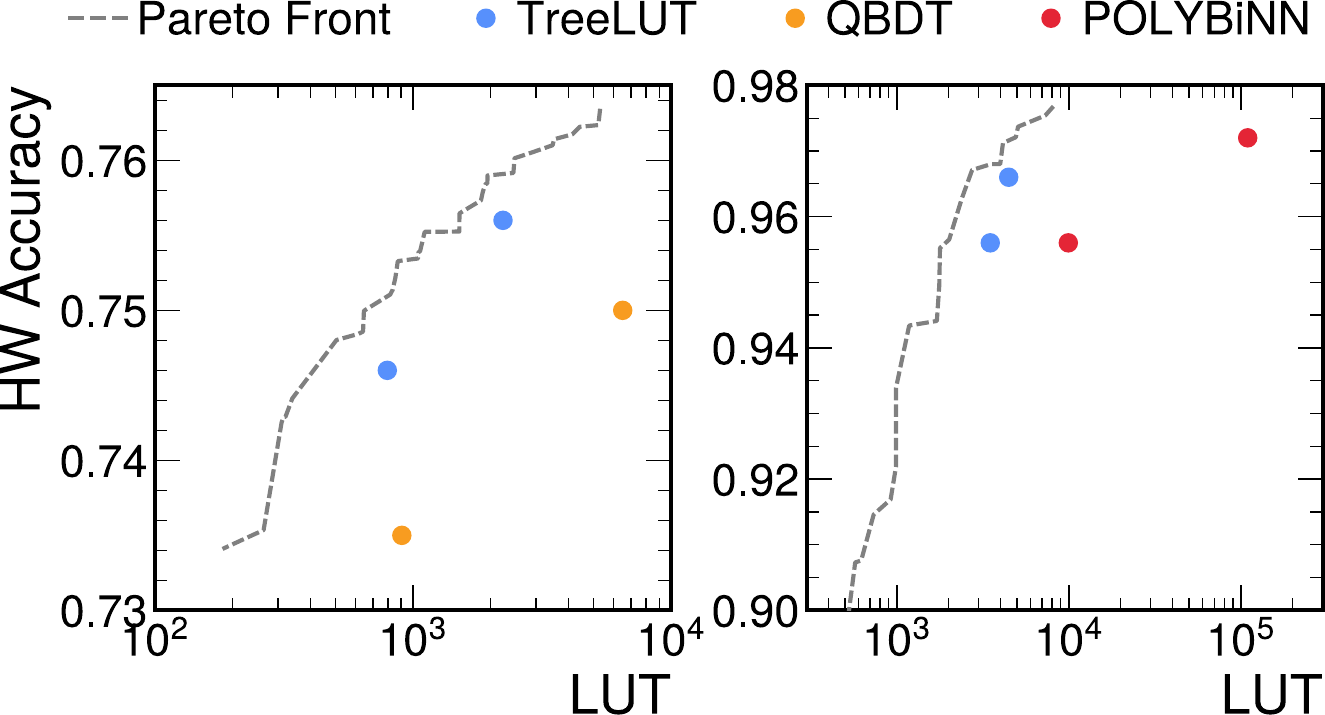}
    \end{center}
    \vspace{-0.3cm}
    \caption{Comparison of FQTree with prior FPGA-based BDT implementations on the JSC (left) and MNIST (right) datasets in terms of accuracy and LUT usage. The Pareto frontier obtained by FQTree outperforms the prior decision-based works across the whole trade-off space.}
    \label{fig:comp}
    \vspace{-0.4cm}
\end{figure}

Figure~\ref{fig:comp} shows a comparison between the Pareto frontier of FQTree and representative prior BDT-based implementations on JSC and MNIST. In both datasets, the FQTree frontier lies above or to the left of the reference points from TreeLUT~\cite{treelut}, QBDT~\cite{ibdt}, and POLYBiNN~\cite{polybinn}, indicating that FQTree can achieve either higher accuracy at a similar LUT budget or lower LUT usage at a comparable accuracy level.

On the NID dataset, FQTree achieves 93.1\% accuracy with only 157 LUTs and 1 cycle (1.9 ns), compared with 92.7\% accuracy, 345 LUTs, and 2 cycles for TreeLUT. This corresponds to about 55\% lower LUT usage together with higher accuracy and lower latency. At smaller operating points, FQTree achieves 92.8\% accuracy with 83 LUTs and 91.7\% accuracy with only 38 LUTs, showing that it can remain highly efficient even under very tight resource budgets. In comparison, QBDT-8bit requires approximately 1,800 LUTs to reach 92\%, while QBDT-1bit uses 170 LUTs for 91.5\%. These results indicate that FQTree is particularly effective at identifying compact high-quality implementations for low-cost BDT deployment.

For completeness, we also evaluate a post-training quantization (PTQ) baseline using exactly the same quantizer as FQTree. The only difference is when quantization is introduced: in PTQ, the BDT is first trained in floating point and the quantizer is applied afterwards, whereas in FQTree the same quantizer is incorporated during training so that the model can adapt to the quantized representation. This comparison separates the benefit of the quantizer itself from the benefit of quantization-aware training optimization. As shown in Table~\ref{tab:bdt}, PTQ already performs very similarly to the state-of-the-art TreeLUT~\cite{treelut} results on both JSC HLF and MNIST, indicating that the underlying quantizer is already competitive for hardware-oriented BDT deployment. FQTree then further improves this baseline by training with quantization awareness, leading to better trade-offs in most cases. For example, on JSC HLF, FQTree improves over PTQ at both reported points, achieving slightly higher accuracy with lower LUT usage. On MNIST, FQTree provides both a higher-accuracy operating point and a more LUT-efficient operating region. The results suggest the gains of FQTree come from both the quantizer's design and the application of quantization during training.

\section{Conclusion}
This work presents a unified workflow for a fine-grained quantization and hardware generation of boosted decision trees targeting low-latency FPGA implementation. By combining quantization-aware training with separate precision control for different numerical components and a compiler-driven hardware flow, our method enables efficient exploration of the accuracy-resource-latency trade-off without manual redesign. Experimental results on JSC, MNIST, and NID show that our method reduces LUT usage by 26-57\% compared to the state-of-the-art TreeLUT while matching or improving accuracy, and outperforms PTQ baselines. 
Future work will extend FQTree to larger BDTs and a broader range of FPGA platforms and applications, 
and explore trustworthiness-aware BDT inference~\cite{que2025trustworthy} and design-flow automation~\cite{metamlpro}.

\vspace{0.1cm}
\noindent \textbf{Acknowledgement.}
Partial support from the United Kingdom EPSRC (grant numbers UKRI256, EP/V028251/1, EP/N031768/1, EP/S030069/1, and EP/X036006/1), and United States DoE (grant numbers DE-SC0011925, DE-FOA-0002705), NSF (grant numbers PHY240298, PHY2117997), and AMD is gratefully acknowledged.

\balance
\footnotesize
\bibliographystyle{IEEEtran}

\bibliography{biblio}

\end{document}